*Polar In-Plane Surface Orientation of a Ferroelectric Nematic Liquid Crystal:*
*Polar Monodomains and Twisted State Electro-Optics*

Xi Chen[1], Eva Korblova[2], Matthew A. Glaser[1], Joseph E. Maclennan[1],
David M. Walba[2], Noel A. Clark[1]*

[1]*Department of Physics and Soft Materials Research Center,*
*University of Colorado, Boulder, CO 80309, USA*

[2]*Department of Chemistry and Soft Materials Research Center,*
*University of Colorado, Boulder, CO 80309, USA*

*Abstract*

We show that surface interactions can vectorially structure the three-dimensional polarization field of a ferroelectric fluid. The contact between a ferroelectric nematic liquid crystal and a surface with in-plane polarity generates a preferred in-plane orientation of the polarization field at that interface. This is a route to the formation of fluid or glassy monodomains of high polarization without the need for electric field poling. For example, unidirectional buffing of polyimide films on planar surfaces to give quadrupolar in-plane anisotropy also induces macroscopic in-plane polar order at the surfaces, enabling the formation of a variety of azimuthal polar director structures in the cell interior, including uniform and twisted states. In a π-twist cell, obtained with antiparallel, unidirectional buffing on opposing surfaces, we demonstrate three distinct modes of ferroelectric nematic electro-optic response: intrinsic, viscosity-limited, field-induced molecular reorientation; field-induced motion of domain walls separating twisted states of opposite chirality; and propagation of polarization reorientation solitons from the cell plates to the cell center upon field reversal. Chirally doped ferroelectric nematics in antiparallel-rubbed cells produce Grandjean textures of helical twist that can be unwound via field-induced polar surface reorientation transitions. Fields required are in the 3 V/mm range, indicating an in-plane polar anchoring energy of $w_P \sim 3 \times 10^{-3}$ J/m$^2$.

*INTRODUCTION*

Nematic liquid crystals (LCs) are useful because of their facile collective response to applied fields and to surface forces [1]. In a liquid crystal, the bulk response is the long-ranged deformation of a fluid, elastic field of molecular orientation, on which confining surfaces establish geometrical and topological structural constraints. In the realm of electro-optics, these two basic elements of LC phenomenology have been combined to create LC display technology [2], thereby enabling the portable computing revolution of the twentieth century [3]. In this development and until very recently, nematic electro-optics has been based on bulk dielectric alignment, in which a quadrupolar coupling to applied electric field induces polarization in a nonpolar LC to generate torque and molecular reorientation. Surface interactions employed to achieve desirable device structures are similarly quadrupolar, with common treatments such as buffing [4] or photo-alignment [5,6,7,8] described by the Rapini-Papoular (RP) model [9] and its variants.

Nematic liquid crystals are fluids having internal long-range orientational ordering. In statistical mechanical terms, because the isotropic-to-nematic phase transition breaks orientational symmetry, it yields Goldstone modes in the nematic, describing spatial variation of the director, $n(r)$, the local average molecular orientation. Because of the full orientational symmetry of the isotropic phase, in the nematic phase the director has no globally preferred orientation and therefore the harmonic (elastic) energy cost of orientational variation of wavevector $q$ decreases to zero as $Kq^2$ at long wavelengths, where $K$ is the orientational (Frank) elastic constant. A bulk nematic sample can be oriented by providing an arbitrarily small force, for example an arbitrarily small applied electric or magnetic field, which couples to nematic orientation via quadrupolar dielectric or diamagnetic anisotropy. The nematic order is similarly infinitely responsive to boundary conditions imposed by surfaces on which there is orientational anisotropy. Because the nematic is a fluid, these conditions make it possible to put a nematic in a container, have it spontaneously anneal into a space-filling, three-dimensional (3D) director orientation structure dictated by the bounding surfaces, and have this structure respond in a predictable way to applied field. In practice, to be useful in applications, the surfaces and fields must be strong enough to eliminate defects and to produce sufficiently fast reorientations of the director.

A novel nematic liquid crystal phase [10,11,12,13] has recently been shown to be a ferroelectric nematic ($N_F$) [14], offering a variety of opportunities to exploit LC field and surface phenomena in new ways. The ferroelectric nematic is a 3D liquid with a macroscopic electric polarization $P(r)$ [14]. On the nanoscale, each molecular dipole is constrained to be nearly parallel to its molecular steric long axis, which translates macroscopically into a strong orientational coupling making $P(r)$ locally parallel to $\pm n(r)$, the local average molecular long axis orienta-tion and uniaxial optic axis of the phase [14]. The polarization thus endows the $N_F$ with coupling between $n(r)$ and applied electric field, $E$, that is linear and is dominant over the dielectric coupling at low $E$. The $N_F$ phase exhibits self-stabilized, spontaneous polar ordering that is nearly complete [12,14], with a polar order parameter, $p = \langle \cos(\beta_i) \rangle \gtrsim 0.9$, where $\beta_i$ is the angle between a typical molecular dipole and the local average polarization density $P$. The resulting large spontaneous polarization ($P \sim 6$ μC/cm$^2$) enables field-induced nematic director reorientation and an associated electro-optic (EO) response with applied fields in typical cells as small as ~1 V/cm, a thousand times smaller than those used to reorient dielectric nematics.

The polar nature of the $N_F$ also results in transformative changes in the interaction of the liquid crystal with bounding surfaces, a key aspect of nematic LC science and its potential for technology. Here we demonstrate that structuring of the vectorial orientation distribution of a 3D volume of polar molecules can be achieved by controlling the polarity of its 2D bounding surfaces. In the simplest example, if the orientation of the preferred polarization is identical on the surfaces of the two parallel glass plates forming a cell, then the $N_F$ volume polarization can be similarly oriented, that is, poled into a uniform orientation by the surfaces, without the need for an applied field.

Materials with spontaneous vectorial order such as ferromagnets and ferroelectrics minimize their energy by breaking up into domains of different orientation of their magnetization, $M(r)$, or polarization, $P(r)$, respectively, with these fields aligned locally parallel to the domain boundaries in order to minimize the energy of the internal and external magnetic or electric fields they produce. Intentionally disrupting this domain structure by applying an external field (*field poling*) is a key process in the use of such materials, *e.g.*, putting an iron rod in a magnetic field to create a bar magnet [15], or actively reversing the polarization of a ferroelectric nanocrystal that serves as a data element in nonvolatile solid-state memories [16]. Field poling of soft materials, such as the corona poling of chromophore-containing polymers to generate poled monodomains for nonlinear optical and electronic electro-optical applications, has been less successful because of the high fields required [17,18]. In ferroelectric nematics, in contrast, we find that suitably polar aligning surfaces can be used to achieve uniform alignment of their nearly saturated bulk polarization, in the absence of an applied electrical poling field.

In the ferroelectric nematic, the reduction in symmetry associated with the appearance of the bulk polarization, $P(r)$, yields a macroscopic coupling of $P$ to applied field $E$, with a corresponding bulk energy density $U_{PE} = -P \cdot E$. In materials where the net molecular dipole is nearly parallel to the steric long axis of the molecule, $P(r)$ is macroscopically constrained to be parallel to $\pm n(r)$ by an orientational energy of the form $U_{Pn} = -u_{Pn}(n \cdot v)^2$, where we define a unit vector polar director $v(r) = P(r)/P$. The energy density coefficient $u_{Pn}$ can be estimated as $u_{Pn} \sim k_B T / vol$ where $vol$ is a molecular volume. If we apply a step change in the orientation of $P(r)$ at $x = 0$, then $n(r)$ will follow within a distance $|x| = \xi \sim \sqrt{(K/u_{Pn})}$, which will be of molecular dimensions. $v(r)$ and $\pm n(r)$ are therefore essentially locked together in orientation on the macroscopic scale, so that if a field is applied, the response is generally to reorient $n(r)$ by reorienting $P(r)$. An exception to this is the movement of pure polarization reversal (PPR) walls, which flip $P$ by 180° with no reorientation of $n$ [14].

## *RESULTS*

*Polar Surface Anchoring* – The intersection of $P(r)$ with a bounding surface deposits polarization charge on the surface, producing electric field-induced, long-ranged interactions with ionic and electronic charge, and with the $N_F$ director [14]. However, bounding surfaces are themselves always inherently polar, since structural variations along the surface normal, $x$, such as the number density or the polarization density of molecular components of the LC or the bounding phase, inevitably lack mirror symmetry about any interfacial plane normal to the surface. Such surface-normal polarity gives polar contributions to the local surface interaction energies of LCs [19,20,21,22,23,24], and in the $N_F$ must be included in $W(\theta,\varphi)$, the local energy/area of $N_F$-surface interaction that depends on $(\theta,\varphi)$, the surface tilt and azimuthal of $P(r)$ about surface normal $x$. A result is that at bounding surfaces, $P_x$ will always be nonzero and will have a preferred sign, so that if the polarization also has an in-plane component, $P$ will be tilted out of the $yz$-plane tangent to the surface. An exception to this would be a situation where the preferred $P_x$ is temperature, $T$, dependent and changes sign at some $T$.

Additionally, the surface structure may have in-plane anisotropy. An RP-type surface treatment, such as bidirectional buffing or normal-incidence photoalignment, creates a surface structure that is azimuthally anisotropic and non-polar with quadrupolar symmetry, meaning reflection symmetric about both the buffing direction and its in-plane normal (*NONPOLAR* case). This quadrupolar symmetry requires, in turn, that $n$ (described by the white, double-headed arrows in ***Fig. 1A***) be *parallel* to the surface. The inherent surface polarity (vertical grey arrows in ***Fig. 1A***) is, in this case, normal to this preferred orientation, and does not induce in-plane, polar symmetry-breaking of $n$. Interaction with a nematic director is given by a surface energy density $W_Q(\varphi)$, where here $\varphi$ gives the in-plane azimuthal orientation between $n$ and the buffing direction, $z$, and $W_Q(\varphi)$ is quadrupolar $[W_Q(\varphi) = W_Q(-\varphi) = W_Q(\pi-\varphi) = W_Q(\pi+\varphi)]$. Following RP, we take $W_Q(\varphi) = w_Q \sin^2\varphi$, where $w_Q$ is a phenomenological coefficient [9,1].

If, on the other hand, in-plane anisotropy is produced by unidirectional buffing along $z$, oblique photoalignment, or some other in-plane, polar treatment in the $(x,z)$ plane, then the surface orientation of a nematic director preferred by this treatment will not be parallel to the surface, but make some (pretilt) angle $0 < \psi < \pi/2$ with the surface plane. Coherent pretilt can be taken as direct indication that such surface treatment, which is procedurally apolar in the sur-face plane, has indeed broken apolar symmetry about the plane normal to the surface and the buffing direction. Such a surface is therefore in-plane polar with a nonzero polar surface energy $W_P$ giving equivalent surface energies for *n* and *-n* in the N phase, but different surface energies for *P* and *-P* in the $N_F$. Such in-plane *POLAR* surface interactions are represented by gray/white arrows in *Fig. 1B*. If $\varphi$ is the in-plane azimuthal angle between *c*, the in-plane component of *n*, and the rubbing direction **z**, then the azimuthal term in the surface energy $W(\varphi)$ must be reflection symmetric about the buffing direction $[W(\varphi) = W(-\varphi)]$, and may be taken to have the form $W(\varphi) = W_Q(\varphi) + W_P(\varphi) = w_Q \sin^2\varphi - \frac{1}{2}w_P\cos(\varphi)$, where $w_P$ is a phenomenological coefficient, as sketched in *Fig. 1B*.

The quadrupolar anisotropy of the LC surface energy density $w_Q$ has been widely studied experimentally with nematics [25]. In the ferroelectric nematic phase, in contrast, $w_P$ is required by symmetry but neither its magnitude nor its sign is known for any particular surface treatment. Here we demonstrate the in-plane polar surface alignment of the $N_F$ phase and obtain an estimate of the polar component of the anchoring energy. We show that such interactions are strong enough to enable full, polar control of the geometry of the nematic director in planar-aligned cells. We create twisted states of the nematic director of the sort pioneered by Mauguin [26] and later used in devices by Schadt and Helfrich [2], and apply in-plane electric fields to produce voltage-controlled optical changes, using much smaller fields and achieving a faster response than in dielectric nematics.

RM734 [10] {Isotropic (I) →182°C → Nematic (N) → 133°C → Ferroelectric Nematic ($N_F$)} was filled into glass cells with a gap of thickness *d* separating the plates, one of which was patterned with ITO or gold electrodes spaced by *L* for application of in-plane electric fields. Cells were studied optically using depolarized transmitted light microscopy (DTLM) with incident white light, and their dynamic response measured using single-wavelength transmission of 632 nm HeNe laser light focused to a 30 µm-diameter spot. Cells both with random-planar glass surfaces, giving a Schlieren texture in the N phase, and with glass plates with buffed polymer alignment layers, giving planar monodomains upon cooling from the isotropic to the N phase, were used.

The polymer-aligned cells included ones with a cell gap *d* = 11 µm and weak, bidirectionally buffed alignment layers that favor alignment of the director of nematic LCs at the surface parallel to the buffing direction and tangent to the surface. With this alignment, because the director (yellow double-arrows in *Fig. 1A*) is perpendicular to the inherent surface-normal polarity, the in-plane apolar symmetry of the director field is not broken (white double-arrows in *Fig. 1A*). These are therefore non-polar, in-plane anchoring surfaces, making a *NONPOLAR* cell.

We also used $d$ = 3.5 μm and 4.6 μm thick cells with unidirectionally buffed polyimide alignment layers that favor alignment of the nematic director parallel to the buffing direction but with a coherent elevation of *n* above the surface plane by a small pretilt angle, $\psi \sim 3°$ [27] (*Fig. 1B*). In this case the apolar in-plane symmetry is broken (gray/white arrows in *Fig. 1B*) and the surface is potentially capable of in-plane polar alignment of the $N_F$ phase, which we refer to simply as being *POLAR*. If the buffing is unidirectional and parallel on the two flat surfaces of a cell, then the surfaces will be termed *SYNPOLAR*, and if the unidirectional buffing is antiparallel they will be termed *ANTIPOLAR*.

Upon cooling a *NONPOLAR* aligned cell with spatially uniform temperature through the N–$N_F$ transition at -1°C/min, we observe a texture of irregular domains extended locally parallel to *n(r)*, first appearing on a submicron scale but then annealing over a roughly 2°C interval into a pattern of lines, of low optical contrast and up to millimeters in length, that are also oriented generally along *n(r)* [14]. These lines coarsen to form closed loops 10–200 microns in extent in some places, that have a distinctive and characteristic lens shape, elongated along *n(r)*, as in *Figs. 1A,C*. These lines form boundaries where there is pure polarization reversal (PPR) [14] between ferroelectric domains with opposite sign of *P* which occupy roughly equal areas of the cell, *i.e.,* show little preference for a particular sign of *P*. Application of a small, in-plane probe field enables visualization of the distinct orientations of *P*, providing prime evidence for the ferroelectric nematic phase in RM734. *NONPOLAR* alignment thus produces quality, uniform alignment of *n* which gives good extinction between crossed polarizer and analyzer.

We also probed the in-plane polar ordering and alignment effects of *single* surfaces in cells using the temperature gradient methods of Aryasova and Reznick [28], who studied the structure and phase behavior of nematics in samples where a constant temperature difference, $\Delta T$, was maintained across the thickness of the cell. Upon cooling through the isotropic-to-nematic phase transition under these conditions, they found that the final nematic alignment obtained was predominantly that favored by the cooler cell surface, concluding that in a temperature gradient, the nematic phase appears first at the cooler surface and the I/N interface then moves as a quasi-planar sheet towards the warmer surface. In a cell with one plate coated with a rubbed polymer film, and the other with an untreated polymer film giving random-planar alignment, this form of cooling was found to produce a uniform, planar-aligned cell upon growing the nematic from the rubbed surface, or, in the same cell, a defected Schlieren texture upon growing the nematic from the random-planar aligning surface.

RM734 was cooled at -1°C/min through the N–$N_F$ transition in *ANTIPOLAR* cells while maintaining a ~2°C temperature difference between the outside of the top and bottom cell plates, with the top plate cooler. The temperature difference across the LC-containing cell gap was estimated to be $\Delta T \sim 0.1°C$ [28,29,30]. In these cells, the N–$N_F$ transition appears as a sharp boundary across which there is a discernably larger birefringence in the $N_F$ phase [14]. As this boundary moves across the cell and toward the lower plate, it leaves behind a polar $N_F$ monodomain, as seen in *Figs. 1D*. During this cooling process, sketched in *Fig. 1B*, the $N_F$ phase nucleates and grows into the nematic from the cooler surface, and, as evidenced by the uniformly orange birefringence color left behind, eventually fills the gap between the plates with a uniformly oriented structure. This state extinguishes between crossed polarizer and analyzer and has a birefringence color indicative of planar or nearly planar alignment, *i.e.*, *n* is uniform through the cell and parallel to the plates. An important feature of such planar-aligned $N_F$ cells with large *P* is their "block" polarization structure [31,32], wherein splay of the ***P-n*** couple and termination of $P_x$ at the cell boundaries is suppressed by the high electrostatic energy cost of polarization charge. Assuming a polarization $P$ = 6 μC/cm$^2$ [14], any surface-imposed pretilt therefore disappears within a polarization self-field penetration length $\xi_P = \sqrt{(K\varepsilon_o/P^2)} \sim 0.1$ nm of the surface, with the ***P(r)-n(r)*** couple rapidly becoming locally parallel to the cell plates, the orientation where $P_x$ = 0. If ***P*** rotates to develop an *x* component, then the $E_x$ resulting from polarization charge at the surfaces acts to return ***P*** to be parallel to the plates. The in-plane orientation of ***P*** can be probed by application of small (few V/cm) in-plane electric fields [14]. The uniform orientation of ***n-P*** in this as-grown monodomain in an *ANTIPOLAR* cell indicates that while the cooler surface nucleates and grows the $N_F$ phase with its preferred orientation of ***P***, the warmer surface is forced to adopt the non-preferred polar orientation. In *SYNPOLAR* cells, both surfaces end up in their preferred states.

If the cooling is halted ~2°C or less below the N–$N_F$ transition, this situation with metastable orientation at the warmer plate persists, with the optically uniform polar monodomain spreading to fill the cell area. However, if the *ANTIPOLAR* cell is cooled further, to ~5°C below $T_c$, a structural transition occurs, with multiple domain boundaries nucleating and moving across the cell, purple-colored regions replacing the uniform orange state, as seen in *Fig. 1D*. These purple regions have their own internal domain walls. At a given place in the cell, the passing of the orange-to-purple boundary mediates a spontaneous transition to macroscopic chirality. Notably, as shown in a thicker cell in *Fig. 2A,B*, optical extinction between crossed polarizers with *n* along the optical polarization is no longer obtained in this new domain, indicating a non-uniform director field which, in the absence of applied electric field, and being constrained only at the surfaces, can only be a twisted state, with *n* parallel to the plates and reorienting helically at a constant $\partial\varphi/\partial x$. Within these twisted regions, internal domain walls separate regions which exhibit identical brightness and color between crossed polarizer and analyzer (*Fig. 1D*) but which become either darker or brighter if the polarizer and analyzer are uncrossed, depending on the sign of the uncrossing angle (*Figs. 2D,E*). This is an optical signature of identical twisted states of opposite handedness, left-handed (LH) or right-handed (RH).

Since in these cells the uniform director field that grows in at the N–$N_F$ transition has the preferred polar orientation on the cooler plate (at $x = d$), and a less energetically favorable condition on the warmer surface (at $x = 0$), we propose that the lower $T$ transition to a twisted state is mediated by a flipping of *n-P* at the warmer surface to its energetically favored orientation, at the cost of introducing a twist elastic energy density in the bulk, $U_T = ½K(\partial\varphi(x)/\partial x)^2$. A convenient parameterization of the anchoring strength of such anisotropic surface interactions is given by the "surface penetration length" $\ell = K/w$, where $K$ is the twist elastic constant and $w$ the relevant energy density coefficient [1]. Applying a torque to the LC director induces a linear $\varphi(x)$, which extrapolates to zero a distance $\ell$ into the surface. Taking $K = 5$ pN and $w_Q \sim 10^{-4}$ J/m² for typical rubbed polyimide alignment [25], we have $\ell$ for the quadrupolar interaction $\ell_Q \sim 50$ nm, so that orientational deformations in the few-micron thick LC films studied here occur with $n(r)$ at the surface essentially fixed along one of the orientations preferred by the quadrupolar anchoring and the fraction of the π twist in the LC sustained by the surfaces is $d/(d + 2\ell_Q) \cong 1$. In other words, the torque transmitted by the bulk Frank elasticity and applied to the surfaces does not pull the orientation at each surface very far from that of the quadrupolar surface energy minima. In a cell with *ANTIPOLAR* surfaces, the minima in $W_Q$ on the warmer surface are at $\varphi = 0$ and $\varphi = \pi$, with the minimum at $\varphi = 0$ being of lower energy, and that at $\varphi = \pi$ of higher energy only because of $W_P$ (*Fig. 1B*) The minimum at $\varphi = \pi$ is therefore metastable, and, as $w_P$ increases in magnitude as $T$ is lowered (due to the increasing magnitude of *P*), a transition from $\varphi = \pi$ to $\varphi = 0$ becomes possible at the warmer surface. For a linear π twist through the cell, the condition for this transition to occur is $w_P > K\pi^2/4d \sim 5\times10^{-6}$ J/m². This minimum required value of $w_P$ is considerably smaller than the typical $w_Q$. In general, the nature of polar ordering of liquid crystals along a surface normal varies widely, ranging from the perfect vectorial orientation of polar molecules, like 5CB on semiconductor surfaces [33,34], to the much smaller induced polarization of mirror symmetric molecules on any surface. Ferroelectric chiral smectics exhibit polar surface interactions, including preferred polarization normal to the surface [35]. A chiral smectic A exhibiting an electroclinic effect was found to have a normal component of polarization at the cell surface experiencing an effective surface field $E_s \sim 2$ V/μm normal to the surface [36]. If the polar force is electrostatic, then the corresponding energy/area, $U_P$, depends on $l$, the thickness of the interaction volume, with $U_P \sim PE_s l$. For the $N_F$ phase, assuming an interaction volume of molecular dimensions, a similar estimate would give $U_P \sim 10^{-4}$ J/m² and $w_P = U_P \sin\psi \sim 10^{-5}$ J/m². Surface polarity can also induce polarization in the LC, leading to flexoelectric and order electric gradients in the vicinity of the surface [37,38,39].

Let us consider in more detail the events at the warmer surface of the ANTIPOLAR gradient cell. As the N–$N_F$ growth front approaches, as in *Fig. 1B*, the $N_F$ phase is replacing the N phase, which contacts the warmer buffed surface with its director lying along one of the quadrupolar minimum energy orientations. The $N_F$ director has the same orientation, but the locally polar orientation induced in the N phase by the surface does not match that of the approaching $N_F$ front: an N-phase gap with opposite signs of *P* on its two surfaces, a kind of PPR wall [14], is thus created and trapped, in a uniform director field. This geometry is at least metastable at temperatures near the N–$N_F$ transition. At lower temperatures, however, the orientation of *n* near the surface changes by π, transitioning to the other quadrupolar minimum so that the bulk matches *P* at the surface.

A possible mechanism for this transition is shown in *Fig. 2J*, where, for the purposes of illustration, we consider the temperature gradient along *x* being reduced, so that the $N_F$ grows in from both surfaces and forms a PPR wall in the bulk. A simple topological transition can take place whereby appropriate twist deformation of *P-n* above and below the PPR wall reduces the magnitude of the component of *P* being reversed in the wall, leading to the replacement of the PPR wall with a twist wall which then spreads along *x* to fill the cell, giving a uniformly twisted πT state. This is a barrier-limited transition because the initial twist must be substantial over the width of the PPR wall. If a gradient in *T* is present, the process of eliminating the PPR wall formed near the warmer surface is essentially the same. Such a barrier-limited topological transition results in the heterogeneous nucleation and growth evident in *Fig. 2*. The sharp boundary between the uniform and π-twisted states is evidence for the metastability of the $\varphi = 0$ surface state at the warmer plate. The surface orientation transition between quadrupolar surface energy minima at $\varphi = 0$ and $\varphi = \pi$ is therefore solitonic in nature, very similar in structure to the surface-stabilized π reorientation walls in chiral smectic ferroelectric LCs (SSFLCs) and similarly described by a double sine-Gordon equation [40].

The topological walls that separate the πT states are 2π-twist (2πT) walls, illustrated in *Fig. 2E*. In the geometry of this cell, with the rubbing directions nearly parallel to the electrode edges, the polarization at mid-height in the cell, $P(x = d/2)$, is normal to the electrodes, in the *y* direction for the left-handed π-twist state, and in the -*y* direction for the right-handed π-twist state, and therefore parallel or antiparallel to the applied field, *E*. The 2πT walls consist of a 2π-twist disclination line in the bulk *n* and *P* fields (*Figs. 2,3*). If we consider just the treated *ANTI-POLAR* cell plates shown *Fig. 2C* and the applied field, ignoring the LC, it is clear that this configuration is chiral, with a handedness that changes with the sign of field. This means that for a given sign of *E*, one of the πT states is of lower energy than the other and therefore that the 2πT walls can be moved by applying an electric field, and will move in opposite directions for different sign of field, discussed further below in the context of *Fig. 3*. The topological transition mediating the field-induced nucleation of the twist wall loops in an *ANTIPOLAR* cell is also barrier-limited, resulting, for example, in the heterogeneous nucleation of LH domains in a RH π–twisted cell shown in *Fig. 3*.

It should be clear from the preceding discussion that virtually any ferroelectric nematic structure in which there is twisted-planar reorientation of the *n-P* couple can be stabilized between plates by using aligning surfaces with bi-directional rubbing or mono-directional rubbing with appropriate pretilt. We will see below that this can be combined with chiral doping to control the relative stability of otherwise degenerate twisted states of different helicity, as in the case of twisted nematic devices [41]. These textures are eventually replaced by that of the crystal phase on further slow cooling (at T ~ 90ºC). Quenching the $N_F$ from T = 120ºC by dunking in room temperature water preserves these and the *SYNPOLAR* textures, discussed next, in glassy form.

In *SYNPOLAR* aligned cells that are cooled homogeneously through the N–$N_F$ transition at -1ºC/min, the well-aligned nematic monodomain observed in the N phase evolves into a uniform, polar monodomain of the $N_F$ phase (*Fig. 2F-I*). These monodomains respond to in-plane fields as small as 1 V/cm applied normal to *P*, as reported in [14] for states of uniform *P*, confirming bulk ferroelectric poling in the absence of an applied electric field.

*In-Plane Electro-Optics and Dynamics* – The field-induced orientational dynamic responses of planar-oriented cells described above, can be classified into two principal types, one characteristic of field rotation (*ROT*), occurring when a component of *E* appears in the direction normal to *P*, and the other typical of field reversal (*REV*), occurring when the component of *E* parallel to *P* changes sign. Assuming that $\varphi(x,t)$ varies only through the thickness of the cell, the bulk response to an in-plane applied field is governed by the dynamic torque balance equation, $\mathcal{T}_E = \gamma \partial\varphi/\partial t = -PE_y \sin\varphi + K\partial^2\varphi/\partial x^2$, where $\gamma$ is the nematic rotational viscosity. Simulations of the field-induced reorientation profiles, $\varphi(x,t)$, with fixed surfaces show these two distinct dynamic limits, ROT in *Fig. 4D* and REV in *Fig. 4B*.

The response of an initially uniformly twisted state to an in-plane field, *E* applied at time *t* = 0 is shown in *Fig. 4D*. This simulation illustrates the *ROT* case, modeling the response of *P* that would follow a step-wise field rotation. Here the applied field favors the orientation $\varphi =0$, the initial orientation of *P* at the cell center, as in *Fig.2*. The final, field-induced polarization state (black curve in *Fig. 4D*) is then subjected to field reversal (the *REV* case). For *NONPOLAR* cells (*Fig. 1A*), the polarization on the surfaces can be reversed independently, producing a mix of left- and right-handed, uniformly twisted π-twist states, and the resulting surface walls can be moved by field-induced torques [14]. In general, then, there may be surface orientation transitions. In contrast, in the 3.5 μm thick *ANTIPOLAR* cell, we do not observe field-induced polar surface reorientations. The surfaces stay in their *POLAR* preferred minima in these *E*-field experiments, and in this is also the case in the simulations presented in *Figs. 4B,D* for an *ANTIPOLAR* cell, which have fixed surface orientations (we will see later that surface transitions can be induced when the sample is chirally doped, as illustrated in *Fig. 5*).

In the following three sections, we consider ROT, domain wall, and REV reorientation. In all cases, the field variable $E$ refers to the magnitude of the field applied at the area of measurement: the probe laser spot in the ROT and REV cases; and in the electrode gap center for the domain dynamics DTLM images. In the gap center $E \approx V_{IN}/L$, where $V_{IN}$ is the voltage applied to the electrode gap, and $L \sim 1.5$ mm is its effective gap width, which is larger than the physical 1 mm gap because of the spread of field in the $x$ direction from the thin edges of the 2D electrodes.

*Field Rotation (ROT) Dynamics* - In the ROT response regime, $\Delta\varphi$, the initial angle between **P** and applied field **E** is smaller in magnitude than 90° ($|\Delta\varphi| \leq 90°$) and the torque applied by the field, $\mathcal{T}_E = -PE\sin\varphi$, tends to rotate **P** toward $\varphi = 0$. In the uniformly twisted starting state of *Fig. 2D*, this condition is obtained everywhere in the cell. Each elements of the cell is effectively elastically decoupled from its neighbors, reorienting independently in the applied field. The polarization responds in a continuous fashion without altering the topology, except for the surface regions at $x = 0$ and $x = d$ where the twist is squeezed into a thickness comparable to the deGennes field penetration length $\xi_E = \sqrt{(K/PE)} \sim 50$ nm for $P = 6$ μC/cm$^2$ and an applied field $E = 40$ V/mm [14]. At fields $E > 1$ V/mm, where the twist is confined to sub-wavelength thick surface regions, the optical transmission of this cell approximates that of a uniform birefringent slab, giving near-extinction with crossed polarizer and analyzer along $y$ and $z$. In the simulations of *Fig. 4*, $E = 40$ V/mm and the surface anchoring is assumed to be infinitely strong, with the top and bottom boundaries fixed at $\varphi = \pi/2$ and $-\pi/2$ respectively.

In *Fig. 4D*, since $\xi_E \ll d$, the ferroelectric and viscous torques are dominant and the Frank elastic term can be dropped from the torque balance equation, yielding $\gamma\partial\varphi(t)/\partial t = -PE_y\sin\varphi(t) = \mathcal{T}_E$. The relaxation function satisfying this equation of motion is given by $\varphi_E(t) = 2\tan^{-1}[\tan(\varphi_o(x)/2)\exp(-tPE/\gamma)]$, where $\varphi_o(x)$ is the starting angle between **P** and $E = V_{IN}/L$. This is a particularly simple dynamic limit in which the ferroelectric/viscous torque balance independently controls the response of each element of length $dx = \xi_E$ of the sample, taking $\varphi(t)$ to be uniform in each such element with a starting value $\varphi_o(x)$ [42]. Thus, *Fig. 4D* also describes the homogeneous reorientation of samples with uniform $\varphi(t)$ with starting orientations in the range $-90° < \varphi_o(x) < 90°$, including $\varphi_o = 45°$ (cyan arrow). These simulations are thus applicable to the dynamical measurements of $\varphi(t)$ shown in *Fig. 4E*. For small $\varphi_o$, the dynamics of $\varphi_E(t)$ in this regime are exponential, with the characteristic relaxation time $\tau_E = \gamma/PE$. With these approximations, $\partial\varphi_E(t)/\partial t \approx \varphi_E(t)/\tau_E$ so that the linearity of the initial twist profile in the interior of the cell should be maintained during reorientation, as is indeed seen in *Fig. 4D*. The ROT reorientational response time $\tau_E$, the intrinsic, viscosity-limited response time of the LC, is the fastest achievable for given $\gamma$ and $P$, scaling as $1/E$ and plotted from the simulation in *Fig. 4D* as the dashed white line in *Fig. 4A*. The electro-optic response in the ROT geometry was measured in RM734 by applying short voltage pulses to the West electrode in the four-electrode geometry of *Fig. 4E* to reorient a 5 µm-thick, random-planar cell at $T$ = 110°C. A small DC bias was initially applied to the electrodes to give $\varphi_o \sim 45°$ and maximize the starting transmitted intensity $I_o$ between crossed polarizer and analyzer. Assuming that $\varphi(x)$ is uniform in $x$, the transmitted intensity is given by $I(\varphi(t)) = I_o\sin^2(2\varphi(t))$. The measured optical response was fit to $I(\varphi_E(t))$ to obtain $\tau_E$ vs. $E$, plotted as the cyan squares in *Fig. 4A*. A typical measured response (solid cyan curve in *Fig. 4E*) fitted by $\varphi_E(t)$ (blue line), yields $\tau_E = \gamma/PE$ = 71 µsec for an $E$ = 40 V/mm field step. In the range in the range 0.64 < $E$ < 36 V/mm, we find that $\tau_E \propto 1/E$, as expected for $\varphi_E(t)$, but for higher fields $\tau_E$ decreases much more slowly with $E$. Fitting the lower field data using $P$ = 6 µC/cm² gives $\gamma$ = 0.15 Pa s, and this viscosity was used in the simulations.

To understand the high-field behavior, we consider that, while the large value of $P$ ensures a strong coupling of orientation to applied field, it also produces ubiquitous space charge effects that may significantly slow the dynamics. In all of the cells with in-plane electrodes studied here, there is an effective thin, insulating barrier at the interface between the $N_F$ phase and the electrodes, a barrier which either has no polarization or in which the polarization is fixed in orientation. Such layers drop part of the applied voltage in maintaining the physical separation of free charge carriers from the polarization charge accumulated at the insulating interface with the ferroelectric LC. Such a layer is formed, for example, by the polymer films used for alignment but even in the four-electrode cell shown in *Fig. 4E*, where there is no alignment layer and the interface is $N_F$/gold, the cell behavior indicates that such an insulating layer exists. If this layer has a capacitance $C$ farads/cm², then the effective field in the above equation for $\varphi_E(t)$ is reduced by the depolarization voltage $(P/C)\cos\varphi_E(t)$, becoming $E = [V_{IN} - (P/C)\cos\varphi_E(t)]/L$. Generally, these interfacial layers are not perfectly insulating, with free charge passing through to the polarization charge layer, in which case they can be modeled as a material of resistance $R$ Ω-cm² in parallel with $C$ (a leaky capacitor). We then obtain for $\varphi(t)$:

$$\varphi(t) = 2\tan^{-1}\{\tan(\varphi_o(x)/2)\exp[-t(P/L\gamma)(V_{IN} - (P/C)\cos\varphi(t)\cdot\exp(-t/RC))]\}. \qquad (1)$$

Solving this equation for the effective relaxation time and varying the parameters $C$ and $RC$ to fit the ROT optical response measurements gives the dotted black curve in *Fig 4A*, with RC = 100 µsec and C = 0.15 µF/cm². This result shows that for low $E$, the response of the director field is sufficiently slow that the depolarization voltage across the barrier relaxes toward zero, allowing the full $V_{IN}/L$ field to appear across the LC material, and yielding $\tau_E = \gamma/PE$ behavior. At intermediate fields, the insulating layer behaves like a capacitor and the applied voltage is divided between the insulating layer and the LC. When the field is applied, substantial reorientation occurs before the boundary charge can leak away, and the depolarization voltage competes with $V_{IN}$ to reduce the effective field on the LC and substantially retard the relaxation. At the highest fields, we have $V_{IN} \gg P/C$, and the relaxation time approaches $1/E$ dependence again.

*Domain-Mediated Dynamics* – Generally, for small, in-plane applied *E*, the LH and RH π-twist states will have different net energies of interaction with the field. As a result, if both twist states are present in a cell, the 2π-twist wall boundaries between them move in the *yz*-plane to expand the area of those states with lower energy, or new lower-energy ones will be nucleated. The creation or motion of these domain boundaries is inherently hysteretic, potentially enabling bistable electro-optic effects. An example of 2π-twist wall creation in a *d* = 3.5 µm *ANTIPOLAR* cell with in-plane ITO electrodes spaced by 1 mm is shown in *Fig. 3*. The field is weak (*E* < 0.3 V/mm) and very slowly varying (~0.1 Hz). The significant, continuous distortion of the original RH πT state with increasing field is punctuated by the nucleation and growth of less distorted LH πT domains, which eventually take over the whole cell area and which remain after *E* is returned to zero. The field direction in this cell is offset by 3° from the bisector of the antiparallel buffing directions, ultimately favoring the global counterclockwise reorientation of *P* (see profile in *Fig. 3J*). As a result, the 2π-twist walls separating the LH domains mutually annihilate on contact and disappear as the LH area grows.

*Field Reversal (REV) Dynamics* – The response to applied field reversal is simulated by starting with the field-confined LH state that is the final configuration of the ROT process (solid black curves in *Fig. 4D*) and flipping the sign of *E*. Immediately after field reversal, the entire center of the cell has *P* opposing the applied field, so that in most of the cell the polarization experiences zero torque from the field and thus does not respond. However, near the surfaces, where *P* reorients to the surface preference, there is an electrical torque so the response starts there, with the formation of solitons in the polarization field (*Fig. 4B*) that propagate toward the cell center at a velocity $v = \xi_E/\tau_E$, reaching the center in the field reversal time $\tau_R = d/2v \approx \tau_E(d/2\xi_E)$ [43]. The solitons then combine to form a thin sheet centered at *x* = *d*/2 where the polarization makes a $\Delta\varphi = 2\pi$ reorientation between equivalent orientations of *P* aligned with the field. This sheet disappears via the topological processes of order reconstruction [44,45], or the spontaneous appearance of twist disclination loops, either way changing the handedness to leave a field-confined RH π-twist configuration. The simulated reversal time $\tau_R$ in the soliton regime, plotted as the upper solid white line in *Fig. 4A*, varies with *E* as $\tau_R \propto E^{-1/2}$, decreasing more slowly with filed than the $\tau_E \propto E^{-1}$ rotational response discussed previously. In our field range, $\tau_R$ is noticeably larger than $\tau_E$, a reflection of the time spent by the metastable interior of the cell with *P* at $\varphi$ = 0 waiting for the solitons to arrive.

Field-reversal dynamics have been probed experimentally by measuring the response of the twisted states in the cell of *Fig. 2* to square-wave driving, as simulated in *Fig. 4B*. The optical response enables identification of the soliton propagation (baby pink) and topological transition (baby blue) stages of reorientation. The optical transmission through crossed polarizer/analyzer (parallel to *E*) vs. time following an applied voltage reversal at *t* = 0 is shown in *Fig.*

*4C*. The transmission of the starting state is very low, with only the very thin, twisted surface regions depolarizing the light. The subsequent response, characterized by one or more peaks in the transmitted intensity, is identical for +/- and -/+ reversals, indicating that a complete transition between an LH and RH confined twist state is obtained at each reversal. As *E* is increased, the overall optical transmission due to the traveling solitons (baby pink arrows) gets smaller because the solitons become thinner [43]. The arrival of the solitons in the cell center is heralded by a distinct peak in the transmission (*Figs. 4B,C*, white circles), produced by the transient, extended linear variation in $\varphi(x)$ about $\varphi(x) = 0$ near the center of the cell (white circles, *Fig. 4B,C*). This peak is directly followed by the order reconstruction that completes the transition to the final field-stabilized, nearly-extinguishing state of opposite handedness (baby blue arrows). Once reorientation is complete, the baseline optical transmission is again very low, with only the confined regions at the cell surfaces depolarizing the light. The overall transition times $\tau_R$ for this process, plotted vs. *E* in *Fig. 4A* (using dot colors that match the curve colors in *Fig. 4C*), are much longer than the measured rotation times $\tau_E$. The simulations of *Fig. 4B* give similar $\tau_R$ values (solid white line) with no adjustable parameters, just using the value of ɣ derived from the ROT data. Also plotted in *Fig. 4A* are the optical and polarization field-reversal 10%-90% response times shown in *Fig. 3C* of Ref. [14], obtained for square-wave driving but in different cell geometries (black dots). These response times are comparable to those of *Figs. 4B,C*, indicative of a retardation similar to that seen in solitonic field reversal. Interestingly, these individual sets of data all exhibit close to a $\tau_R \propto E^{-1}$ dependence rather than the $\tau_R \propto E^{-1/2}$ expected from the soliton model. This discrepancy may be due to the assumption in the model that in the beginning, away from the surfaces, there is perfect polar order, with *P* initially uniformly aligned by the field at $\varphi = 0$. In this idealized situation, when the field reverses, each soliton must travel half the cell thickness *d*/2 in order to switch the cell. In practice, however, due to defects, surface imperfections, and thermal fluctuations, *P* is only rarely perfectly aligned along $\varphi = 0$ in the cell interior, and, once the field switches, experiences torques in many different locations and begins to reorient inhomogeneously. This process will proceed more rapidly as the field is increased, and will lead to the local nucleation and soliton-mediated growth of substantially reoriented domains, depending on the size and orientational deviation of nucleated domains [46]. To the extent that such domains appear and expand, the net time for complete reversal is expected to decrease because the solitons will, on average, have less distance to cover. The experimentally observed $\tau_R \propto E^{-1}$ behavior would imply a mean soliton travel distance decreasing with increasing *E* as $E^{-1/2}$. The field-reversal data in *Fig. 4A* suggest that this may be universal behavior.

*Chiral Doping of Ferroelectric Nematic* – We have demonstrated that it is possible to manipulate $N_F$ twisted states by chiral doping. To this end, we carried out electro-optic experiments on a chiral ferroelectric nematic ($N_F^*$) phase, a mixture of RM734 with the widely-used chiral dopant

CB15* [47,48,49] at a concentration $c$ = 10 wt%. RM734 and CB15* mix in the I and N phases, but submicron-scale spots appear at the N to $N_F$ transition, indicating that CB15* may phase separate to some extent. Given the helical twisting power $h \sim$ 5 to 6 (µm wt%) of CB15* in standard nematic hosts [47,48], we would expect the chiral nematic pitch at a concentration $c$ wt% to be in the range 1.6 µm < $p \sim h/c$ < 2.0 µm. In the experiments reported below, we estimate the pitch in the $N_F$ phase to be $p \sim$ 2.3 µm, at the high end of the expected range.

The chiral mixture was filled into $d$ = 3.5 µm thick *ANTIPOLAR* cells and subjected to in-plane fields in the range 0 < $E$ < 8 V/mm. The textures and corresponding cell structures observed in DTLM are shown in *Fig. 5.* We observe a field-induced helix unwinding, mediated for positive $E$ by surface transitions between preferred and metastable surface states (*Fig. 1B*), and for negative $E$ by the bulk 2π twist disclinations of *Fig 2*. The cell geometry is that of *Fig. 2C*, so that, as discussed above, if $E$ is nonzero the cell structure without liquid crystal is chiral, and changes handedness if the sign of $E$ changes. Therefore, since the RM734/CB15* mixture is chiral and of fixed handedness, its response to positive and negative fields should be different, as is clearly observed in the experiments.

*Positive Field Response* – Optical changes are observed in an increasing positive $E$ field, starting at $E \sim$ 0.04 V/mm as seen in *Fig. 3*. As the applied field is increased, we observe the sequential appearance of three topologically distinct states, the grey-to-yellow *initial* (*i*) state, the intermediate orange states (*b,t*), and the maroon-blue final, high-field state for positive $E$ (*f+*), evidenced by the passage of sharp domain walls, as shown in *Fig. 5A-E*. This observed sequence is interpreted under the assumption that, due to polarization charge stabilization, $P$ in the bulk is everywhere in the *yz*-plane, as in the preceding discussions. The final, high-field *f+* state evolves continuously from the maroon hue of *Fig. 5E* to the simulated blue color of *Fig. 3*, indicative of a field-induced uniform orientation in the cell interior with a π/2 twist near each cell surface, as in *Figs. 3I, 4D*. The $E$ = 0 *initial* (*i*) state has the lowest effective birefringence of the sequence, implying that it is the most twisted (by some multiple of π because of the *ANTIPOLAR* surfaces), meaning that this sequence of states corresponds to a field-induced unwinding of the director twist, as might be expected. A key feature of this sequence is that there are two distinct orange states, (*b - bottom*) and (*t -top*), that evolve from the *initial* state *i* via domain wall-mediated transitions. The *b* and *t* states have identical color and have the property that if pairs of *bb* or *tt* states come together, they can merge with no trace of a boundary. In contrast, *bt* domain pairs appear to overlap kaleidoscope-like to give the *final* state (*bt → f+*), without physical interaction, but becoming a maroon color in the overlap region. Since the maroon *final* state is uniform (apart from a π/2 twist near each surface), the *i* to *b* and *t* and then to *f+* transformations must be topological transitions representing distinct but structurally equivalent expulsions of twist from the *i* state. These expulsions, when overlapped, give the *f+* state. Furthermore, because the *b,t* states have the same color, the cell must otherwise be uniform, meaning that if we let *i, b, t*, and *f+* represent the net twist $\Delta\varphi$ in each state, we must have *i - b - t = f+*, or *i = 2b + π*, which, if we take *b,t* = π, the minimal angular jump for a topological transition in an *ANTIPOLAR* cell, gives *i* = 3π. The sequence *i → b or t → f+* is then a sequence of states where the net azimuthal change $\Delta\varphi$ between the two surfaces evolves with increasing *E* field as 3π → 2π → π, with an *E* = 0 chiral pitch in the *initial* 3πT state of *p* = 2.3 µm (2*d*/3). This value should be the bulk pitch, within a ±π/2 reorientation at one surface. The polar director states corresponding to this sequence and the transitions between them are illustrated in Fig. *5F-J*. Of note is that the reductions in $\Delta\varphi$ at the *i → b* (or *i → t*) events are mediated by *surface* orientational transitions from the global minimum in $W(\varphi)$ (the preferred surface polarization orientation) to the metastable minimum in $W(\varphi)$, sketched in the inset of *Fig. 1B*.

DTLM images of the $\Delta\varphi$ = 3π *initial* state obtained with polarizer and analyzer uncrossed in opposite directions and compared with simulated spectral colors, as in *Figs. 3I-K* confirm that CB15* in the RM734 N* and $N_F$* phases is a right-handed nematic dopant, inducing a right-handed helix as it does in 5CB.

*Negative Field Response* – For negative *E*, the principal transition out of the *initial* 3πT state is to the field-induced *f-* state shown in *Fig. 5Q*. This sign of field expands the partially aligned π-twist regions near the surfaces, leaving the surface orientations unchanged, and squeezing the central 2π twist into a thin metastable sheet, which is then eliminated by the heterogeneous nucleation of holes ( 2π dislocation loops) that spread and eventually fill the central plane, leaving the *f-* state. In the *f-* state, the field enforces uniform orientation in the cell interior, there are π/2 twist walls near each cell surface, and the surfaces remain at their preferred orientations. Note that this is structurally equivalent to the *f+* state but rotated by π, so that the two surfaces are in their preferred states, making the *f-* state the lowest energy field-induced state in the chirally doped cell. Both the *f+* and *f-* states are stabilized by the field, but they, and the *b,t* states, all relax in distinctly different ways when the field is reduced or removed. If the field is removed from the *f-* state, the director field finds itself extremely dilated relative to the *initial* $\Delta\varphi$ = 3π state, and return to the *i* state by the nucleation and passage of 2π bulk twist disclination, like that shown in *Fig. 2E.*

*Relaxation on Field Reduction* –The behavior observed upon reducing the *positive* field back to *E* = 0 is sketched in *Figs. 5K-N*. The $\Delta\varphi$ = 2π and 3π states relax, with the compressed boundaries of the *b* and *t* structures of the 2π-twist state expanding into the formerly uniform interior (*Figs. M,N,D,E*). The field-stabilized *f+* state relaxes to a linear $\Delta\varphi$ = π twist (magenta arrow), which, like the *f-* , is extremely dilated relative to the *initial* $\Delta\varphi$ = 3π state, but, unlike the *f-*, can increase the twist by changing its surface states, reorienting $\varphi(0)$ and $\varphi(d)$ from their metastable minima back to the preferred minima, as sketched in *Fig. 5L*. One such surface transition generates a 2π–twist state. If *E* is reduced slowly, this process is observed as the motion of domain walls that expand the areas of neighboring 2π-twist states. However if *E* is reduced more quickly (in just a few seconds), a striking instability appears that produces the modulated textures shown in *Figs. 5A,K,L*. This modulation is possible because the surface reorientations of π can take place on either the top or the bottom surface, giving RH 2π-twist states of diametrically opposing *P(x)* (*Fig. 5L*). These form a spatially periodic pattern of field-free 2π-twist states, where in each half-period there is one preferred and one metastable surface orientation, alternating between top and bottom between half-periods. The origin of this instability is not clear, but spontaneous periodicity is not uncommon in viscous dynamic LC systems due to the coupling of flow and reorientation in response to conditions of strong driving [50,51].

Once formed upon reducing *E*, these modulated areas can return to the *initial* state at *E* ~ 0 by a process in which the modulation stripes appear to split apart into an array of distinct lines separated by areas which have reverted to configurations with the surfaces aligned along the preferred polar minima, *i.e.*, to the *initial* Δ$\varphi$ = 3π state, resulting in textures of homogeneous *i* areas and irregular remnant lines, as in *Figs. 5A,O*. These lines are the $N_F^*$ analog of oily-streak defects in an N* texture, but, as is the case for all of the $N_F$ phase textures, have *n(r)* and *P(r)* in the *xy*-plane. The structure of these remnant lines at *E* = 0, sketched in *Fig. 5P*, reflects various mixes of 2π metastable states with one polar, non-preferred surface orientation, or the 3π *reverse* state (*r*) with both surfaces metastable. The *i* state interspersed with these defect lines is the final condition following a positive-*E* cycle starting and ending at zero. Applying a *negative E* field to this modulated starting state splits the defect lines into two walls, with their internal 2πT areas appearing in between. For *E* < 0, the 2πT areas immediately transform into the *f*-state via a surface transition to the favored orientation at the bottom of the cell that relieves the twist there, as illustrated in *Fig. 5Q*. In the absence of defect lines, the *f*- state can also be reached from the *i* state via the formation and disappearance of a 2π disclination sheet in the cell interior (*Fig. 5Q*), as discussed previously. The *E* ~ 3V/mm range of the field required to obtain the *f* states enables an estimate of the energy difference between the preferred and non-preferred polar surface orientations, by taking a maximum to be $w_P = PE\xi_E = \sqrt{(PEK)} = $ ~ $3 \times 10^{-3}$ J/m². By comparison, the typical quadrupolar wells of rubbed polyimide films are in the range $10^{-4}$ J/m² < $w_Q$ < $10^{-2}$ J/m².

## *Discussion*

*Nematic and Ferroelectric Nematic Textures* – Liquid crystals feature a remarkable combination of fluidity and order, which manifests itself in their uniquely facile macroscopic response to surfaces. In thin cells between flat plates, the bulk LC organization is strongly affected by surface interactions, leading to the formation of specific textures, bulk structural themes on the micron-to-millimeter scale that reflect the balance of surface, field, and elastic energies. Analysis of these textures, typically using depolarized transmitted light microscopy, enables the study both of surface alignment characteristics and of key bulk LC properties like phase behavior, elasticity and symmetry. Friedel's discovery of smectics is the classic example of this investigative approach [52]. If the orientational preferences on the two surfaces of a cell are different, then generally each preference should be manifest somehow in the bulk, either in any given cell, or as a statistical average over the LC organization in many cells. For example, in a hybrid-aligned cell, where one side prefers *n* normal to the surface and the other *n* parallel, the observed splay-bend nematic texture simultaneously satisfies both surface preferences. If the two surfaces are equivalent in structure but differ in preferred orientation (rubbed in perpendicular directions, for example), then the different surface preferences must be expressed equivalently in the bulk LC in some fashion, as they indeed are in the 90º twisted texture that forms in a nematic. A planar-aligned smectic, on the other hand, may accommodate the surfaces in such a 90º-rubbed cell by forming homogeneous domains that locally fill the cell, with some domains following the 0º surface and others following the 90º surface, with areas that are equal when averaged over many cells.

This brings us to cells of the $N_F$ phase, in which, new to LC alignment physics, a similar condition appears for the *polar* ordering at the surfaces. If the surfaces are structurally identical but differ in polar orientation, then the competing polar orientations must be equivalently expressed in the cell texture. The first definitive examples of such behavior in nematics are the πT states of opposite handedness in the *ANTIPOLAR* $N_F$ cells of ***Figs. 1,2***. These twisted states have distributions in azimuthal orientation that are symmetric about $\varphi = \pi/2$, thus accommodating the competing polar surface orientations of the two cell surfaces. This scenario is achieved by growth in a temperature gradient of the $N_F$ from one surface leading to the formation of a PPR wall at the other surface which becomes unstable at lower temperature, leading in turn to the formation of a uniformly twisted state. Cooling without a gradient leads to PPR formation in the cell interior, as in ***Figs. 1F,2H***, giving a uniform director field with a PPR wall parallel to the cell plates that effects a transition between the two surface preferences. This kind of growth could lead to the formation of arrays of twist domains with different surface preference. The transition to the twisted state in ***Fig. 2C*** is barrier-limited so the PPR wall may be stable or metastable, depending on its energy cost relative to that of the uniformly twisted state. When cooling without a gradient, N/$N_F$ interfaces move into the cell from both surfaces and may interact with one another. For example, thermal interactions could lead to a coupled Mullins-Sekerka instability [1], as the approach of one interface releasing heat will slow the advance of the other.

*Solvent Poling* – The bulk ferroelectric nematic LC can be thought of as a polar medium described by the continuous unit vector field $v(r) = P(r)/P$, the 3D structure of which is determined by the Frank elasticity, the electrostatic self-energy, and interactions with its bounding surfaces.

Such a 3D polar medium can, in turn, act as a vectorial solvent, locally orienting polar solute molecules along $v(r)$. This would constitute poling *by* the solvent, in contrast to the typical approach of field poling *in* a solvent (e.g., in a polymer thermoset) [53]. Ferroelectric nematics are extremely unusual solvents owing to their fluidity, high degree of polar orientational order, and record-large, liquid-state ferroelectric polarization. The lack of mirror reflection symmetry along ***P*** guarantees some degree of induced polarity in any solute molecule, but dipolar solutes, in particular, immersed in the RM734 $N_F$ phase find themselves in a highly polarizing environment. The pair correlation functions from our atomistic simulations [14] show that the mean $E$ field experienced by each molecule due to its neighbors in the $N_F$ phase is of the order $E_{loc} \sim 10^9$ V/m, parallel to ***P***. This is effectively a poling field, creating a $\sim 10\, k_BT$ barrier that orientationally pins the dipoles of individual RM734 molecules, or of single, similarly dipolar solute molecules in RM734, parallel to ***P*** on average. Thus, for molecules with appropriate steric and dipolar structure (such as rod-like molecules with large, longitudinal dipole moments), $v(r)$ is effectively a strongly aligning field ultimately controlled by the orientational preferences of the surfaces.

In summary, we have shown that in the ferroelectric nematic, surface polarity takes on an entirely new role in providing coupling to a macroscopic field variable. Polarity normal to the bounding surfaces is inherent, but must be brought under control to be useful in $N_F$ materials. In-plane polarity will result from any surface treatment, *e.g.*, coating, deposition, etching, or illumination, that locally breaks mirror symmetry about some plane normal to the surface. Finally, in order for studies of the optical textures or other behavior of ferroelectric nematics to be interpretable, the polar alignment characteristics of the bounding surfaces must be specified, understood, and reported.

## *Materials and Methods*

***Synthesis of RM734*** – 4-[(4-nitrophenoxy)carbonyl]phenyl 2,4-dimethoxybenzoate (RM734) is a rod-shaped mesogen first synthesized by Mandle et al. [10]. It was reported to have an isotropic (I) phase and two additional phases with nematic character, with transition temperatures as follows: I – 187°C – N – 133°C – $N_X$ – X. Our preparation gives transition temperatures as follows: I – 188° C – N – 133°C – $N_F$ – X.

***Observations of Response to Applied Electric Field -*** Experimental cells were made by filling LC samples between glass plates spaced to a desired gap, *d*. The plates were coated with lithographically patterned ITO or gold electrodes for application of in-plane electric fields. Cells with unidirectionally buffed alignment layers were obtained from Instec, Inc. Electro-optic observations were carried out with cells mounted on the rotary stage of a polarizing light micro-scope and imaged in transmission. Temperature control was maintained using an Instec HCS402 hot stage.

*Acknowledgements*

This work was supported by NSF Condensed Matter Physics Grants DMR 1710711 and DMR 2005170, and by Materials Research Science and Engineering Center (MRSEC) Grant DMR 1420736.

*Figures*

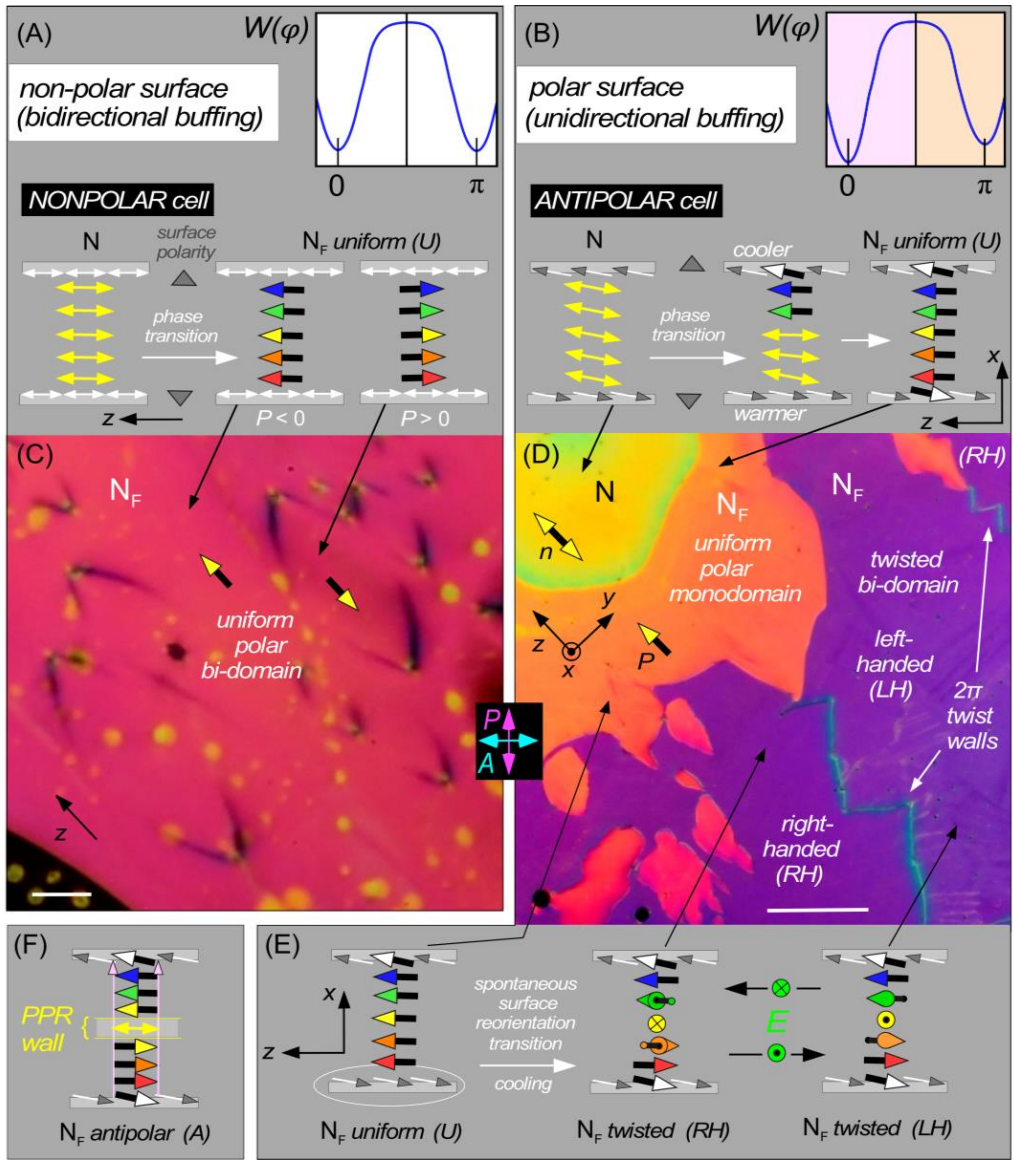

*Figure 1*: Cell structures of the $N_F$ phase of RM734 under different planar alignment conditions, imaged using depolarized light transmission microscopy. (**A,C**) A *NONPOLAR* cell with bidirectional surface buffing gives planar alignment and no pretilt (director parallel to the surface). The cell surfaces are non-polar, allowing the formation of domains of opposite ferroelectric polarization ($d$ = 11 μm, $T$ = 110°C). The plot in (A) shows the corresponding orientational surface energy, $W(\varphi)$, having deep quadrupolar wells along the rubbing direction. (**B,D,E**) A cell with unidirectional surface buffing gives *POLAR* planar alignment with pretilt of a few degrees ($d$ = 4.6 μm, $T$ ~130°C). The in-plane anchoring of $n,P$ at each surface is polar, because the unidirectionality of the buffing breaks mirror symmetry normal to the surface plane and buffing direction, as evidenced by the coherent pretilt. The plot of $W(\varphi)$ then shows a stable preferred well, shaded in pink, and a metastable well, shaded in orange, differing in $\varphi$ by π. Cool-

ing through the N–N$_F$ transition with a temperature difference between the plates grows a polar-oriented N$_F$ monodomain from the cooler plate (sketched in B, orange region in D). This cell is *ANTIPOLAR*, due to the antiparallel buffing on the two plates. As the cell is cooled further into the N$_F$ phase the orientation near the warmer plate eventually undergoes a surface reorientation transition to its lower-energy polar state, creating a π twist in the *n-P* field (E). This twist can be either left-handed (LH) or right-handed (RH), with 2π-twist walls separating these two states. The LH and RH states (purple) are optically degenerate when viewed between crossed polarizer and analyzer. (**F**) Schematic diagram of an antipolar (A), uniform director state, obtained when the cell is cooled homogeneously, the polar N$_F$ phase growing in independently from both surfaces and eventually forming a pure polarization reversal (PPR) wall where the two domains meet in the interior. Scale bars: (C) 20 µm, (D) 200 µm.

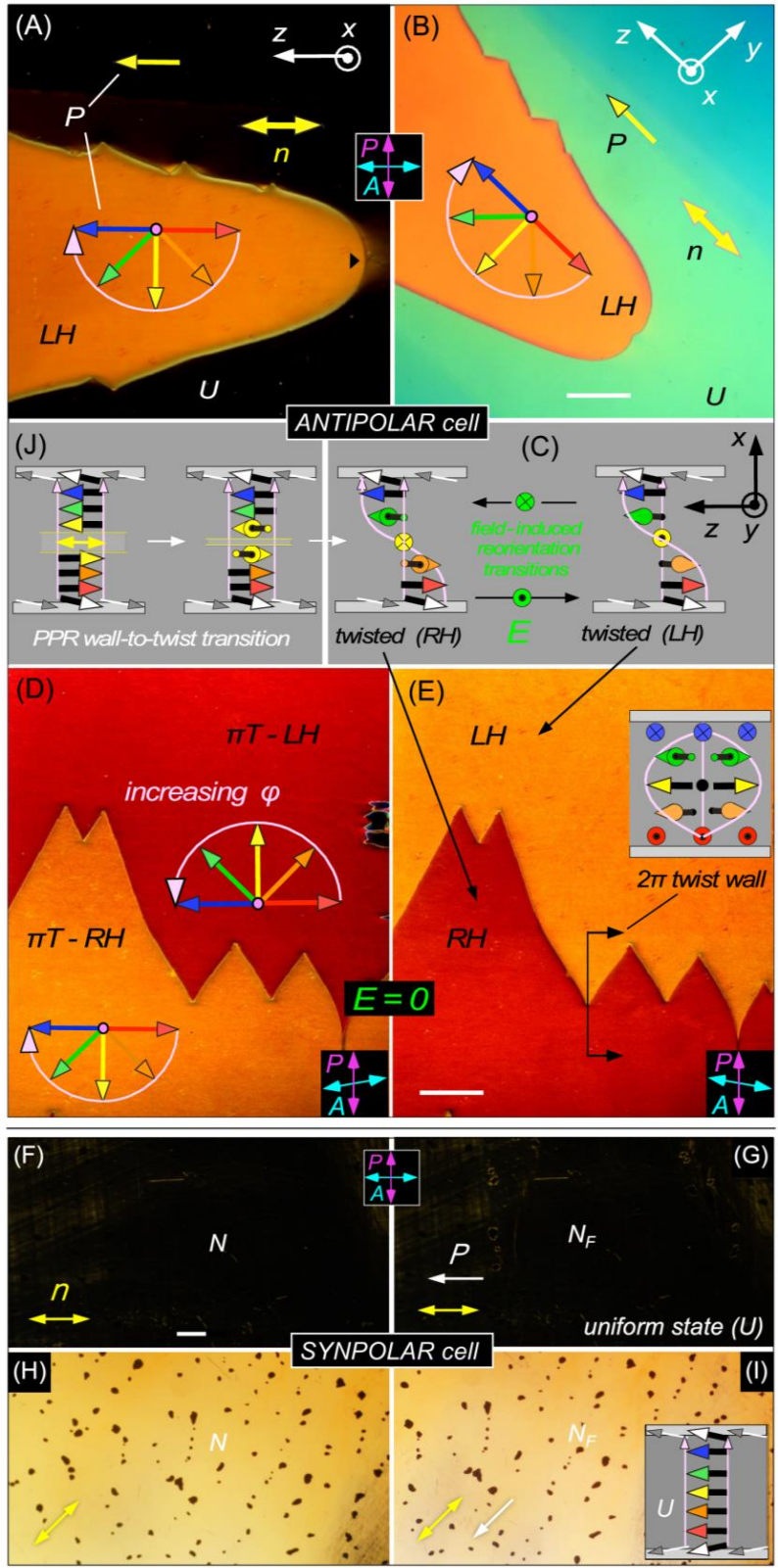

*Figure 2*: Orientational states of *ANTIPOLAR* and *SYNPOLAR* cells. Stable states in the $N_F$ phase induced by polar surface anchoring generated using unidirectional buffing, imaged using DTLM on cooling from T = 135°C to *T* = 125°C. The red-to-blue arrow sequence represents the ferroelectric polarization orientation at increasing heights in the cell (x = 0, *d*/4, *d*/2, 3*d*/4, *d*, where *d* = 3.5 μm in A-E and 15 μm in F-I). The pink arrows indicate the azimuthal angular trajectory of $\varphi(x)$ from the bottom to the top of the cell, which is also the light propagation direction. (**A,B**) A *d* = 3.5 μm *ANTIPOLAR* cell with a left-handed (LH) πT state (orange) growing into a uniform (U) state. The U state is dark between crossed polarizer and analyzer when *n* is either parallel or normal to the analyzer, but shows light green/blue birefringence when rotated (**B**). The color of the πT state does not depend strongly on cell orientation. (**C**) Schematic diagram of the LH and RH πT states. In an *ANTIPOLAR* cell with the field applied, *E* and the white polarization vectors at the surfaces form a triad which is structurally chiral and changes handedness if *E* changes sign, implying that field reversal will tend to flip the handedness of the *n,P* structure. (**D,E**) Decrossing the analyzer lifts the optical degeneracy of the LH and RH states, revealing their chirality and optical symmetry under simultaneous mirror reflection and reversal of the decrossing angle. The LH and RH states are separated by a 2π-twist wall. (**F-I**) Uniform (U) $N_F$ state obtained on cooling a *d* = 15 μm *SYNPOLAR* cell from the N phase. With *n* parallel to the crossed analyzer the cell is dark (F,G), showing quality, planar alignment of the director in both phases except near air bubbles (dark spots) in both phases. In the $N_F$ phase, the polarization is uniformly aligned, with no domains of opposite polarization observed anywhere in the cell. The sample is rotated through 45° in (H,I). (**J**) Spontaneous transition of a pure polarization reversal (PPR) wall to a twisted state. In the absence of a gradient in T, the uniform director state has formed and grown in independently from each surface, making a PPR wall near the cell center. The twist deformation at the PPR wall (yellow vectors) initially costs local twist Frank energy but the system eventually lowers the energy by effecting a topological transition that converts the PPR wall into uniform director twist between the cell plates. This transition is barrier-limited. For cooling an *ANTIPOLAR* cell in a temperature gradient, the PPR wall forms near the warmer surface. Scale bars: 200 μm.

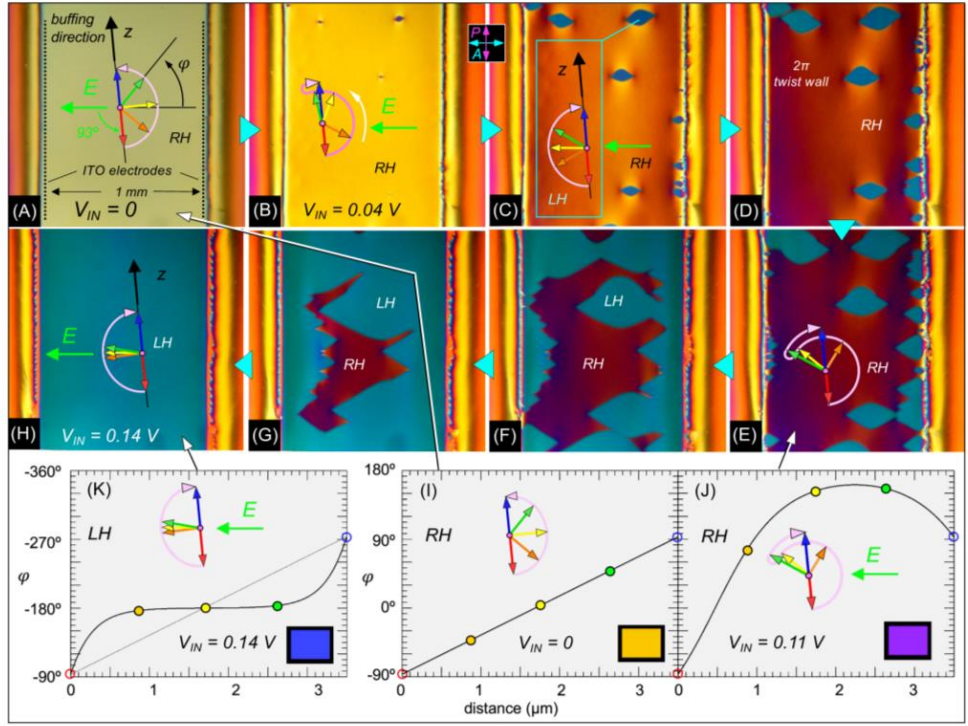

*Figure 3*: Low-field polarization reversal in an *ANTIPOLAR* cell of RM734 at $T$ = 125°C. The cell thickness is $d$ = 3.5 μm. In-plane ITO electrodes with a 1 mm gap are used to apply an electric field $E$ normal to the electrode edges. At the gap center, $E = (V_{IN}/L)$ V/mm, where $V_{IN}$ is the voltage applied to the electrodes and $L \sim 1.5$ mm their effective spacing due to the thin-electrode field geometry. The buffing directions are along the z axis, which is oriented 3° from the electrode edges and therefore 93° from $E$. (**A**) Initial right-handed (RH) πT state in the absence of field. (**B-H**) Gradually increasing the applied voltage in the range $0 < V_{IN} < 0.4$ V distorts the RH twist state and leads to the nucleation and growth of LH twist regions. The red-to-blue arrow sequences represent the ferroelectric polarization orientation at increasing heights in the cell (x = 0, $d/4$, $d/2$, $3d/4$, $d$). The pink arrows indicate the azimuthal angular trajectory from the bottom to the top of the cell, which is also the light propagation direction. The 3° offset between the buffing and electrode orientations breaks the mirror symmetry about $E$, causing the *n-P* couple to rotate preferentially counterclockwise in response to $P \times E$ torques (white arrow in (B)). At $V_{IN} \sim 90$ mV ($E \sim 60$ mV/mm in the gap center), the LH πT state appears in several places via heterogeneous nucleation, being the field preferred state since it has $P$ largely directed along $E$. The field tends to expel the LH twist to the surfaces, filling the cell center with the preferred orientation. The LT and RH πT states are separated by 2π-twist walls. The domain walls in (E-H) move readily in response to small increases in the applied field but the internal structure of the LH and RH states changes little in this voltage range. (**I-K**) Steady-state director profiles $\varphi(x)$ of the LH and RH states calculated numerically by solving the field/elastic torque balance equation given in the text, assuming fixed surface orientations at $\varphi(0) = -87°$ and $\varphi(d) = 93°$. (I) Uniformly twisted RH starting state in the absence of applied field. (J) Deformed RH state in presence of

-25-

an E field favoring $\varphi = \pm180°$. Because the buffing is not perpendicular to the applied field, and *P* in the starting state in (I) has a component directed opposite *E*, there is a net counterclockwise reorientation of *P* in the cell. (K) The field finally induces a transformation, shown schematically in *Fig. 1E*, from the RH to the LH twist state. The polarization profile, corresponding to the azimuthal angular trajectory sketched in (H), is plotted here with $\varphi$ increasing downward. Insets show the calculated transmitted hue of the model twist states between crossed polarizer and analyzer.

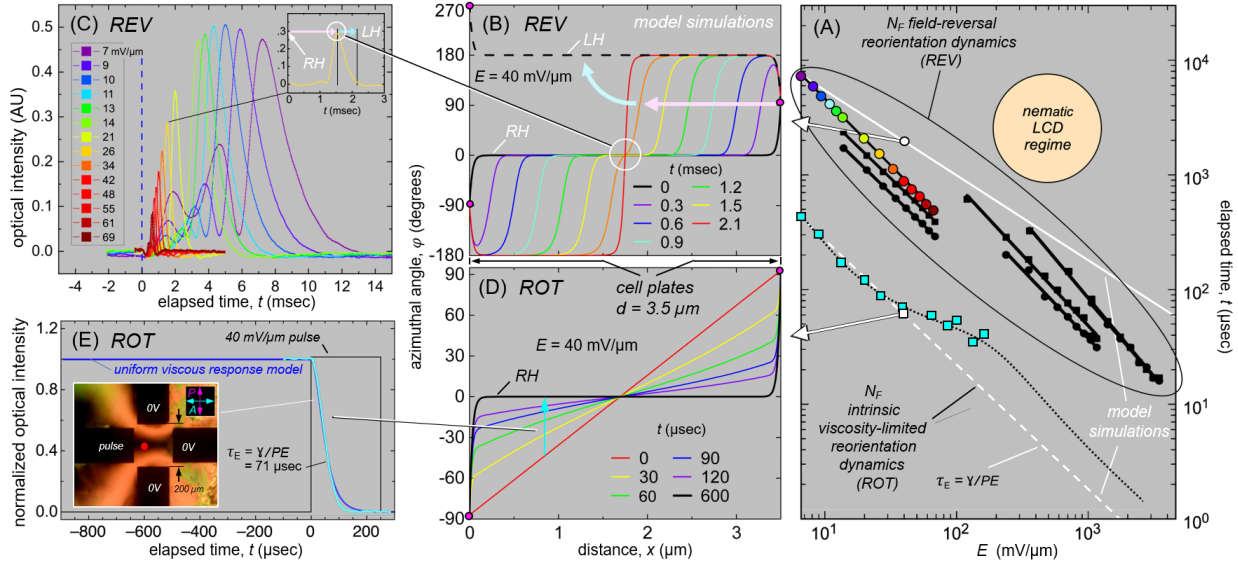

*Figure 4*: Electro-optics and dynamics of RM734 in the $N_F$ phase at $T = 110°C$. (**A**) Experimental and model field-induced polarization reversal (REV) and viscosity-limited (ROT) reorientation times. REV regime: black points are field reversal data from Ref [14]; colored filled circles are peak positions from (C); white circle and solid white line give soliton arrival times at the cell center in (B). ROT regime: cyan squares are reorientation times from curve fits of $\varphi_E(t)$ obtained by fitting the transmission data in (E); white square and dashed white line correspond to reorientation of the type shown in (D); dotted black curve is the leaky capacitive interface model (Eq. 1). REV times are longer because field reversal generates some degree of solitonic response as in (B), in which parts of the cell wait for a soliton to pass in order to reorient. (**B**) Simulated response of the polarization to an electric field reversal (REV). The cell is initially in a distorted RH πT state (black curve). When the field is flipped, *P* in most of the cell finds itself in a state of low-torque, unstable equilibrium ($\varphi \approx 0$), where it remains until solitons, which form at the surfaces, pass by. The final state (red curve) has a 2π-twist wall at the cell center, which disappears by order reconstruction (cyan arrow), leaving the LH πT state (dashed black curve). Arrival of the solitons at the cell center (white circle) produces local director reorientation that gives the optical transmission peaks in (C). (**C**) Optical response between crossed polarizer and analyzer following field reversal. Peak transmission times are plotted in (A) using the same color coding. A similar optical response is observed for the either sign of field reversal, showing that each reversal completely switches the handedness of the πT state. (**D**) Simulated ROT response of a uniformly twisted RH πT state to an applied E field favoring its mid-cell orientation at $\varphi = 0$. In the final field-induced state (black), the field penetration length of the surface orientation into the cell, $\xi_E = \sqrt{(K/PE)}$, is small. In this limit, each element $dx = \xi_E$ of the director profile $\varphi(x)$ responds independently to the field, as $\varphi_E(t)$, so that this graphic effectively also shows the time course of $\varphi(t)$ in ROT reversal for different starting $\varphi_o$ (see text), and can be used to analyze the uniform $\varphi(t)$ field rotation experiment in (E). (**E**) Four gold-electrode cell (NSEW) with random-planar alignment. A small DC bias sets the starting orientation $\varphi_o$ to ~45°. A pulsed voltage (40 mV/μm for 250 μs) is then applied to W, with NS & E grounded, and the transmission $I(t)$ between crossed P and A is measured. 10% to 90% response times are shown vs. field amplitude as the cyan squares in (A). In weak applied fields, the response times decrease as $1/E$ and are used to determine $\gamma$. The deviation from $1/E$ dependence at intermediate $E$ is an effect of depolarization voltage at the electrodes, modeled as a leaky capacitor (black dots, see text).

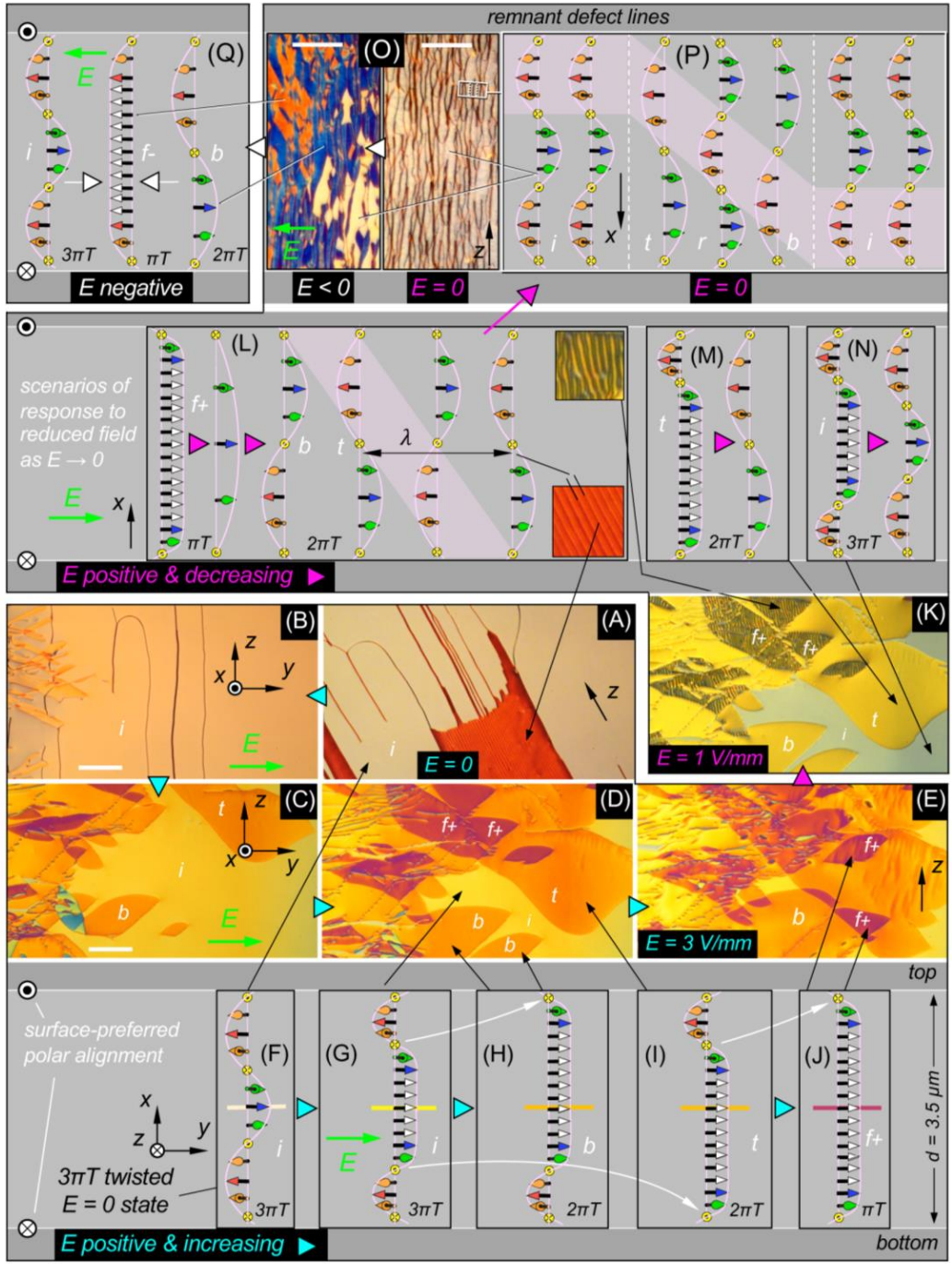

*Figure 5*: Electric field-induced helix unwinding of the $N_F^*$ phase of a chiral RM734/CB15* mixture in a $d$ = 3.5 µm thick *ANTIPOLAR* cell. The white dots/crosses on the left of the gray panels indicate the in-plane polar orientation preference of the surfaces, corresponding to the gray/white arrows in *Fig. 1*. The observed helical structures are all right handed. The empty cell structure with the field applied is chiral, with a handedness depending on the sign of the field. The cyan, magenta, and white triangles indicate the direction of increasing positive, decreasing positive, and increasing negative field, respectively. (**A-E**)

DTLM textures with increasing positive $E$ in the range $0 < E < 3$ V/mm.  (**F-J**)  Sketches of the local 1D profile of $P(x)$ in the *initial* (*i*), *bottom* (*b*)/*top* (*t*), and *final* (*f+*) states, with $P$ becoming more uniform as the field is increased.  The curved white arrows in this panel indicate surface reorientation transitions, where π twist in the interior of the cell are eliminated by surface reorientation from a stable to a metastable state.  The *initial* state (F) is a 3π twist with a pitch $p \sim 2.3$ μm.  Applied field expands the stable region in the middle of the cell (G) and then induces a transition to the *b* and *t* states, which have field-confined π-twists near the cell bottom (H) and top (I), respectively.  Further increase of the field induces a transition to the *f+* state, which has a field-confined π/2 twist on each surface, shown in (E).  (**K**) Upon reducing the applied field, within a few seconds the *b, t* and *f+* domains relax and their boundaries retract to their original locations in (D).  The *f+* domains exhibit a dramatic instability, breaking up into stripes, remnants of which are also shown in (A).  (**L**) Schematic drawing of the *f+* domain instability following a reduction in the applied field.  The initial *f+* relaxation is to a state of uniform π twist with a pitch three times that of the *i* state and a metastable polar surface orientation on both surfaces.  This dilated state relaxes in turn by flipping the orientation at one surface by π, to generate a 2π-twist state.  A periodic instability occurs along the *y* direction when such flips occur alternately on the top and bottom surfaces, generating a metastable array of 2π-twist states differing in average orientation by π.  (**M,N**) Continuous relaxation of the field-distorted *t* and *i* states when $E \to 0$.  The pink shaded areas in (P) and (L) highlight regions of common orientation.  (**O,P**) The system relaxes further from the 2π-twist states to the homogeneous 3π–twist starting condition, *i*, by relaxation of the remaining metastable surface orientations.  A splitting of the periodic array in (L) occurs, with the surface-preferred 3π-twist state appearing between remnant, oily-streak-like defect lines.  These lines are single- or few-period runs of the original surface-metastable 2π or reverse 3π-twist (*r*) state, as shown in (**P**).  Application of a weak negative field further splits the defect lines into two walls, with the interior metastable 2π states appearing in between.  These immediately transform into the *f-* state.  (**Q**) Since the handedness of the empty cell/field structure changes with the sign of field while that of the $N_F^*$ phase is fixed by the dopant, a negative $E$ must produce a completely different sequence of states.  The transformation from *i* to *f-* is achieved either by way of a 2π disclination sheet in the cell center, with fixed surfaces, or from a 2π state via the remnant defect lines of *Figs. 5O,P*.  The *f-* is the lowest energy field-on state, because the surfaces are both at their preferred orientations.  Scale bars: 100 μm